\documentclass[twocolumn,showpacs,preprintnumbers,amsmath,amssymb,prb]{revtex4}
\usepackage{graphicx}
\usepackage{dcolumn}
\usepackage{color}
\usepackage{bm}
\usepackage{float}
\usepackage{gensymb}
\begin{document}

\title{Engineering the magnetic   properties of the Mn$_{13}$ cluster by doping}

\author{Soumendu Datta,$^1$ Mukul Kabir,$^2$ Abhijit Mookerjee$^1$ and Tanusri Saha-Dasgupta$^1$}

\affiliation{$^1$ Advanced Materials Research Unit and Department of Material Sciences, S.N. Bose National Centre for Basic Sciences, JD Block, Sector-III, Salt Lake
City, Kolkata 700 098, India \\
$^2$Department of Materials Science and Engineering, Massachusetts Institute of Technology, Cambridge, Massachusetts 02139, USA }

\date{\today}

\begin{abstract}
With a goal to produce giant magnetic moment in Mn$_{13}$ cluster which will be useful for practical applications, we
 have considered the structure and magnetic properties of pure Mn$_{13}$ cluster and substitutionally doped it with  X = Ti, V, Cr, Fe, Co, Ni atom to produce Mn$_{12}$X clusters. We find that Ti and V substitutions in Mn$_{13}$ cluster are the most promising in terms of gaining substantial binding energy as well as achieving higher magnetic moment through ferromagnetic alignment of atom-centered magnetic moments. This has been demonstrated in terms of energetics and electronic properties of the clusters. For comparison, we have also studied the effect of N-capping of Mn$_{13}$ cluster, predicted in the earlier work [Phys. Rev. Lett. {\bf 89}, 185504 (2002)] as a means to produce stable giant magnetic moment in Mn clusters upto cluster size of 5 Mn atoms.
\end{abstract}
\maketitle

\section{\label{sec:intro}Introduction}

Behavior of Manganese is very special both in the bulk and cluster forms. The electronic configuration of the atom has a half filled 3$d$ shell and filled 4$s$ shell. Structural and magnetic properties of manganese clusters have been studied in the past.\cite{ref1,ref2,pederson,boba1,boba2,smallmnt1,smallmnt2} It has been shown \cite{ref1,ref2,pederson,boba1,boba2} that small Mn-clusters containing up to 4-5 atoms exhibit ferromagnetic coupling of individual atomic magnetic moments with a magnetic moment per atom comparable with the magnetic moment of Mn at the atomic limit, which is 5 $\mu_B$. Interestingly, the coupling becomes ferrimagnetic as the cluster size increases and the net magnetic moment
 falls abruptly as shown in the experimental works \cite{ref3,expt} and also supported by the first principles studies.\cite{ref4} The atom centered individual magnetic moment, though, remains of the order of atomic moment. Therefore, it will be of great technological interest if the ferrimagnetic coupling can
 be transformed into ferromagnetic, thereby resulting into a giant magnetic moment. One way to accomplish this is by doping.\cite{doping} Furthermore, if the doping also increases the stability of the cluster compared to the pure Mn-cluster 
of same size, then it will be an additional achievement on top of {\it spin engineering}. Such clusters will be promising building block for making new cluster assembled materials. Along the same line of thought, Rao and Jena \cite{ref5} previously studied the structure and magnetism of N-capped Mn clusters. Based on their study carried out on Mn clusters of size upto 5 Mn atoms, they proposed that N-capping can be used as a viable means to enhance both the stability and the ferromagnetic coupling among Mn atoms in a Mn-cluster. In the present work, we have proposed an alternative route and studied systematically the effect of substitutional doping of Mn$_{13}$ cluster to produce Mn$_{12}$X clusters with X = Ti, V, Cr, Fe, Co, Ni atom. We find that Ti and V substitutions are the most effective as both the magnetic moment and stability increase significantly for the Mn$_{12}$Ti and Mn$_{12}$V clusters compared to the pure
 Mn$_{13}$ cluster. To compare our finding with the proposal of Rao and Jena\cite{ref5} through N-capping, we have also studied the structure and electronic properties of Mn$_{13}$N cluster. Although, the N-capping enhances the stability and magnetic moment compared to the pure Mn$_{13}$ cluster, the gain in total magnetic moment is found to be smaller compared to that of Mn$_{12}$Ti and Mn$_{12}$V clusters. Also the substitutional Ti-doping and V-doping are more interesting
 since they keep the volume of the cluster intact as opposed to N-capping which increases the volume of the original clusters. The rest of the paper is organized in the following manner. Section II describes the computational details. Section III summarizes the results obtained in our study. This section consists of five sub-sections. The sub-section A is devoted to the results describing pure Mn$_{13}$ cluster. The sub-sections B and C are devoted to results on Mn$_{12}$X clusters. In sub-section D, we analyze the origin of enhanced stability and magnetic moments in Mn$_{12}$X clusters. Sub-section E describes our comparative study on N capped Mn$_{13}$ cluster. We end the paper with a conclusion in section IV.

\section{\label{sec:methodology} Computational Details}

Our calculations have been performed using density functional theory, within the pseudo-potential plane wave method as implemented in the Vienna {\it ab initio} simulation package.\cite{kresse2} We have used the projector augmented wave pseudo-potentials \cite{blochl,kresse} and Perdew-Bruke-Ernzerhof exchange-correlation functional \cite{perdew} for spin-polarized generalized gradient approximation. The wave functions were expanded in the plane wave basis set with the kinetic energy cut-off of 270 eV and reciprocal
space integrations were carried out at the $\Gamma$ point. Symmetry unrestricted geometry optimization has been performed using the 
conjugate gradient and the quasi-Newtonian methods until all the force
components became less than a threshold value of 0.005 eV/{\AA}. Though the noncollinear effect of atomic spins is significant for Mn-clusters, it has been
 shown in the earlier work \cite{noncollinear} that for Mn$_{13}$ cluster, the degree of noncollinearity is very small. Therefore, we have optimized the structures
 for all the spin multiplicities $M = 2S+1$ under the approximation of {\it collinear} atomic spins arrangement to determine the magnetic moment of the minimum energy structure. Simple
cubic super-cells were used with the periodic boundary conditions,
where two neighboring clusters were kept separated by at least 12 {\AA}
vacuum space to make the interaction between the cluster images
negligible.
The binding energy ($E_B$) of each Mn$_{12}$X cluster has been calculated with respect to the free atoms as
\begin{equation}
E_B({\rm {Mn}}_{12}{\rm X}) = 12E({\rm Mn}) + E({\rm X}) - E({\rm Mn}_{12}{\rm X})
\end{equation}
where $E$(Mn$_{12}$X), $E$(Mn) and $E$(X) are the total energies of the Mn$_{12}$X cluster, an isolated Mn atom
 and an isolated $X$ atom respectively.
We also have analyzed the gap between the highest occupied molecular orbital (HOMO) and the lowest unoccupied molecular orbital (LUMO) for the 
 magnetic clusters by defining the spin gap as follows:
\begin{center}
\begin{eqnarray}
\delta_1 = - \left[\epsilon^{\rm majority}_{\rm HOMO} - \epsilon^{\rm minority}_{\rm LUMO}\right]\nonumber\\
\delta_2 = - \left[\epsilon^{\rm minority}_{\rm HOMO} - \epsilon^{\rm majority}_{\rm LUMO}\right]
\label{spingap}
\end{eqnarray}
\end{center}
The system is considered to be stable if both $\delta_1$ and $\delta_2$ are positive. Detailed discussion about $\delta_1$ and $\delta_2$ can be found elsewhere.\cite{ref6,new1,new2,new3}

\section{Results and Discussions}
\subsection{Pure Mn$_{13}$ cluster}
Since the transition metal clusters generally prefer compact geometry to maximize the interaction between the rather localized 
$d-d$ orbitals,\cite{alonso} we have considered icosahedral, cube-octahedral and hexagonal closed pack (HCP) structures as the most probable
 starting geometries to determine the minimum energy structure (MES) for Mn$_{13}$ cluster. These three are the most compact and highly
 coordinated structures for the 13 atoms cluster. Each structure was allowed to relax for {\it all possible} collinear spin configurations
 of the atoms. Fig.1 shows the variation of total binding energy as a function of total magnetic moments for these three structures of Mn$_{13}$ cluster.
\begin{figure}
\begin{center}
\rotatebox{0}{\includegraphics[height=5.5cm,keepaspectratio]{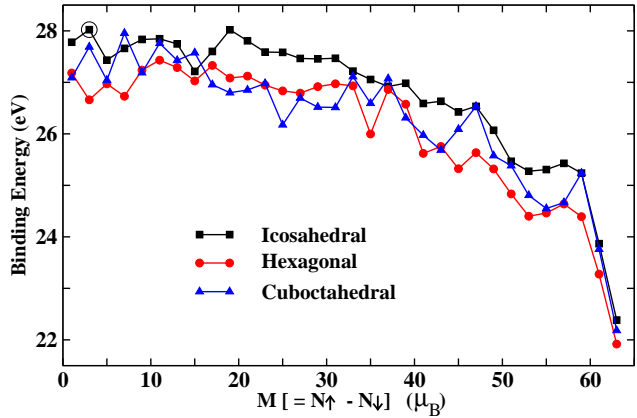}}
\caption{(Color online) Plot of total binding energies for icosahedral, hexagonal closed pack and cuboctahedral structures of the Mn$_{13}$ cluster as a function of total magnetic moment (M) which is the difference between the number of electrons in the up ({\it i.e} N$\uparrow$) and down ({\it i.e} N$\downarrow$) spin components. The data point corresponding to the minimum energy Mn$_{13}$ cluster has 
been encircled.}
\label{mn13}
\end{center}
\end{figure}

 We found an icosahedral structure with total magnetic moment of 3 $\mu_B$ and total binding energy of 28.023 eV is the MES for the Mn$_{13}$ cluster, as predicted
 earlier.\cite{ref4} An icosahedron of 13 atoms can be viewed as consisting of two pentagonal rings connected with each other through the central atom and capped with two apex atoms at its ends as shown in Fig. 2. We also observed a spin segregated state of atomic moments in the MES of Mn$_{13}$ cluster in the sense that the atoms within each of the two pentagonal rings of icosahedral Mn$_{13}$ cluster are ferromagnetically aligned, while the two pentagonal rings are anti-ferromagnetically coupled with each other (cf. Fig.2). This is in agreement with the previous report by another group.\cite{smallmnt2}

\begin{figure}
\begin{center}
\rotatebox{0}{\includegraphics[height=4.5cm,keepaspectratio]{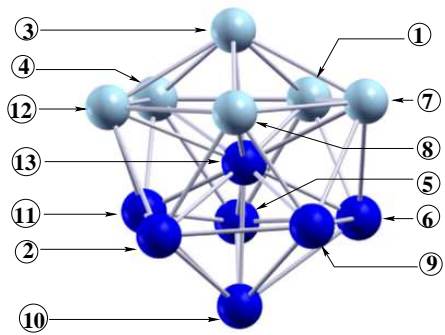}}
\caption{(Color online) Atomic spin configurations for the minimum energy structure of pure Mn$_{13}$ cluster. The cyan (light gray) colored balls correspond 
to atoms with positive magnetic moment and blue (dark gray) colored balls correspond to atoms with negative magnetic moment. The convention for numbering the atoms is shown, which is used in Fig.7.}
\label{mn13}
\end{center}
\end{figure}

\begin{figure*}[]
\begin{center}
\rotatebox{0}{\includegraphics[height=5.6cm,keepaspectratio]{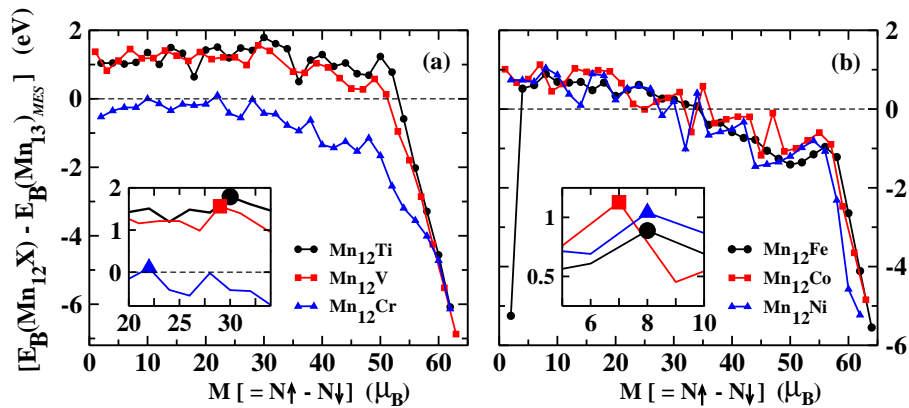}}
\caption{(Color online) Plot of total binding energies of icosahedral Mn$_{12}$X clusters as a function of total magnetic moment(M). The total binding energies are plotted with respect to the binding energy of the MES of icosahedral Mn$_{13}$ cluster. Panels (a) and (b) correspond to the substitutions by elements left and right of Mn in the periodic table respectively. The dotted lines through zero in panels (a) and (b) correspond to the binding energy of MES for pure Mn$_{13}$ cluster. The data points around the minimum energy structure of each Mn$_{12}$X cluster have been emphasized in the insets with the MES by specially marked.}
\label{mn12x}
\end{center}
\end{figure*}

\subsection{Energetics of Mn$_{12}$X clusters}

For the calculations reported in the following, we have considered doped Mn$_{12}$X clusters of the icosahedral symmetry with the impurity $X$ atom sitting at the center of icosahedron and optimized it for {\it all possible} collinear spin configurations of atoms. We have considered six different substitutions at the central Mn site of Mn$_{13}$ cluster: considering the substituting $X$ atom as one of the three elements left of Mn in the periodic table (Ti, V, Cr) and of the three elements right of Mn in the periodic table (Fe, Co, Ni). This assumed geometry of icosahedral symmetry with the impurity $X$ atom sitting at the center of icosahedron  needs some discussion.  In our earlier work on V-doped Co clusters,\cite{ref6} we have also shown by ab-initio molecular dynamic study that the ground state structure of Co$_{12}$V cluster, adopts an icosahedral pattern with the substituting V-atom always preferring to sit at the central site of icosahedral Co$_{12}$V cluster. Theoretical works on clusters with isosahedral symmetry with substitution at the central site have been reported for MCo$_{12}$ (M=Ti, V, Cr, Mn, Fe, Co and Ni) clusters,\cite{new4} Fe$_{12}$X (X = Be, O, Mg, Al, Si, S, Ti, Mn, Fe, Ni, Zn, Se, Pd) clusters,\cite{new5} and Al$_{12}$X (X=B, Al, Ga, C, Si, Ge, Ti, As) clusters\cite{vkumar}. Therefore, there exists enough reports in the literature considering the substitution at the central site by atoms of similar or larger sizes, justifying 
our choice of the impurity $X$-atom doped at the central site of icosahedral Mn$_{13}$ cluster.
Furthermore, as discussed later in section III.D,  Mn-X dimers were found to have higher binding energies
compared to Mn$_2$ dimer. This in turn would indicate that single atom substitution of Mn by X atom
will prefer a configuration where the number of Mn-X bonds are maximized, which is possible for
the substitution at the central position. Our total energy calculations by putting the substituting
atoms at different sites of the icosahedral Mn$_13$ cluster support this idea.

Fig.3 shows the plot of binding energies as a function of total magnetic moments (N$\uparrow$ - N$\downarrow$) of all the Mn$_{12}$X clusters. In order to see the effect of substitutional doping distinctly, we have plotted the binding energy for each spin multiplet of Mn$_{12}$X clusters with respect to that of the MES for pure Mn$_{13}$ cluster {\it i.e} E$_B$(Mn$_{12}$X)- E$_B$(Mn$_{13}$)$_{MES}$. Positive value of this quantity means gain in binding energy upon substitutional doping. It is seen from Fig.3(a) that gains in binding energies are achieved in case of Ti and V substitutions for a range of values of total magnetic moments, from $\sim$ 2 $\mu_B$ to $\sim$50 $\mu_B$, while for Cr-substitution there is no gain in binding energy except a small gain for the magnetic moments of 22 $\mu_B$ and 10 $\mu_B$. On the other hand, for Fe, Co and Ni substitutions, there is also gain in binding energy for each of them upto a total magnetic moment of around 30 $\mu_B$ as seen from Fig.3(b). Notice that these gains are not as high as we observed in case of Ti and V substitutions. However, they are better than the Cr-substitution.

\subsection{Energetics and magnetic moments of MESs of Mn$_{12}$X clusters}

\begin{table}

\begin{center}
\caption{\label{table1}Total binging energies and total magnetic moments (M) of the minimum energy icosahedral structures of Mn$_{12}$X clusters, X = Ti, V, Cr, Mn, Fe, Co, Ni.}
\begin{tabular}{cccccc|ccccc}
\hline
\hline
Clusters && Binding   &&  M     && Clusters && Binding   && M   \\
         && energy    &&        &&          && energy    &&      \\          
         &&  (eV)     && ($\mu_B$)&&          &&  (eV)     && ($\mu_B$)\\
\hline
            &&        &&        &&              &&             &&     \\
Mn$_{12}$Ti && 29.812 && 30     &&  Mn$_{12}$Fe && 28.911      & & 8       \\
Mn$_{12}$V  && 29.582 && 29     &&  Mn$_{12}$Co && 29.151      & & 7        \\
Mn$_{12}$Cr && 28.122 && 22     &&  Mn$_{12}$Ni && 29.062      & & 8       \\
Mn$_{13}$   && 28.023 &&  3     &&              &&             & &          \\
\hline
\end{tabular}
\end{center}
\end{table}

\begin{figure*}[t]
\begin{center}
\rotatebox{0}{\includegraphics[height=4.4cm,keepaspectratio]{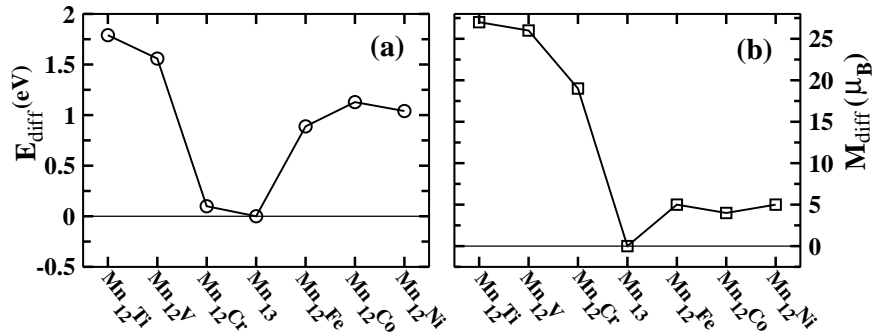}}
\caption{(a) Binding energies [E$_{diff}$ = E$_B$(Mn$_{12}$X)$_{MES}$ - E$_B$(Mn$_{13}$)$_{MES}$] and (b) magnetic moments [M$_{diff}$ = M(Mn$_{12}$X)$_{MES}$ - M(Mn$_{13}$)$_{MES}$] of the minimum energy structures of Mn$_{12}$X clusters with respect to
 those of the minimum energy pure Mn$_{13}$ cluster respectively. The straight lines through zero in panels (a) and (b) are the reference fixed at the
binding energy and total magnetic moment respectively of the MES of pure Mn$_{13}$ cluster. }
\label{diff}
\end{center}
\end{figure*}

Table \ref{table1} shows the total binding energies and total magnetic moments of the MESs of Mn$_{12}$X clusters. It is seen that the MESs of all the doped clusters have higher binding energies and magnetic moments as compared to those of the MES for Mn$_{13}$ cluster. We note again that the MESs of the Mn$_{12}$Ti and Mn$_{12}$V clusters have significantly higher binding energies than Mn$_{13}$. Similarly, the total magnetic moments are found to be significantly large for the MESs of clusters doped with elements to the left of Mn in the periodic table. Moreover, there are several isomers [as seen from Fig. 3(a)] of Mn$_{12}$Ti and Mn$_{12}$V clusters with total magnetic moment in the range of 3-50 $\mu_B$, which have gain in binding energies compared to the MES of pure Mn$_{13}$ cluster. Combining the results of Fig.3 and Table 1, it is therefore seen that the substitutional Ti-doping as well as V-doping of the Mn$_{13}$ cluster are the most preferable in view of gaining significant amount of both the binding energy and the total magnetic moment. For the MES of Mn$_{12}$Cr cluster, although the gain in magnetic moment is quite significant, the gain in binding energy is the least. Similarly, Mn$_{12}$Fe, Mn$_{12}$Co and Mn$_{12}$Ni clusters also have some isomers with total magnetic moments in the range of around 3-30 $\mu_B$ [as seen from Fig. 3(b)], which have both gains in binding energy and magnetic moment with respect to those of the MES for pure Mn$_{13}$ cluster.  However, the MESs for Mn$_{12}$Fe, Mn$_{12}$Co and Mn$_{12}$Ni clusters have total magnetic moments of 8 $\mu_B$, 7 $\mu_B$ and 8 $\mu_B$ respectively, which are higher than that of the MES for Mn$_{13}$ cluster too, but significantly lower than that of the MESs for Mn$_{12}$Ti and Mn$_{12}$V clusters. In order to have a visual depiction of the trend in the choice of substituting elements in a concise manner, the binding energies and magnetic moments of the MESs of Mn$_{12}$X clusters with respect to those of the minimum energy pure Mn$_{13}$ cluster respectively {\it i.e} E$_B$(Mn$_{12}$X)$_{MES}$ - E$_B$(Mn$_{13}$)$_{MES}$ and M(Mn$_{12}$X)$_{MES}$ - M(Mn$_{13}$)$_{MES}$ are plotted in Fig.\ref{diff}.

\begin{figure}[t]
\begin{center}
\rotatebox{0}{\includegraphics[height=10.5cm,keepaspectratio]{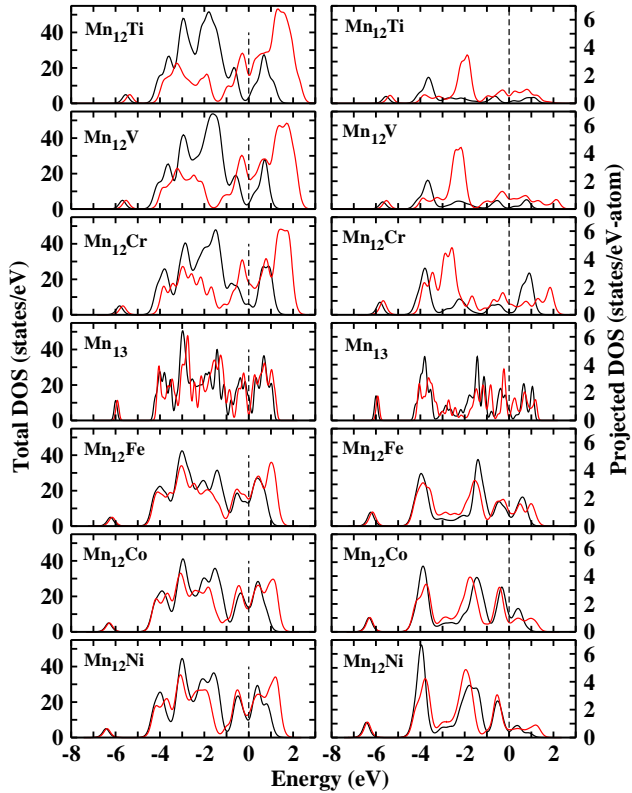}}
\caption{(Color online) Plot of total density of states (left panel) and center atom projected total density of states (right panel) of the minimum energy icosahedral structures of Mn$_{12}$X clusters. Black curves correspond to the results for majority spin channel and red dotted curves for minority spin channel. The dotted line through zero parallel to y-axis in each system corresponds to Fermi-energy of that system. Gaussian smearing (0.1 eV) has been used.}
\label{dos}
\end{center}
\end{figure}

In Fig.\ref{dos}, we have plotted the total density of states and center atom projected density of states of the MES of each Mn$_{12}$X cluster. As is seen from Fig.\ref{dos}, the difference between the majority and minority spin states is large for the optimal icosahedral Mn$_{12}$Ti, Mn$_{12}$V and Mn$_{12}$Cr clusters, which results in large magnetic moments for these clusters. On the other hand, the magnetic moments for the optimal icosahedral structures of Mn$_{13}$, Mn$_{12}$Fe, Mn$_{12}$Co, Mn$_{12}$Ni clusters are relatively small, as the majority and minority spin states mostly cancel each other. Center-atom projected density of states shows that it has finite but small magnetic moment in case of MESs of Mn$_{12}$Ti, Mn$_{12}$V and Mn$_{12}$Cr clusters. On the other hand, its magnetic moment is vanishingly small for the MESs of Mn$_{13}$, Mn$_{12}$Fe, Mn$_{12}$Co and Mn$_{12}$Ni clusters.

\subsection{Origin of stability and enhanced magnetism of the MESs of Mn$_{12}$X clusters}
\begin{figure}[b]
\begin{center}
\vskip 2.0cm
\rotatebox{0}{\includegraphics[height=4.1cm,keepaspectratio]{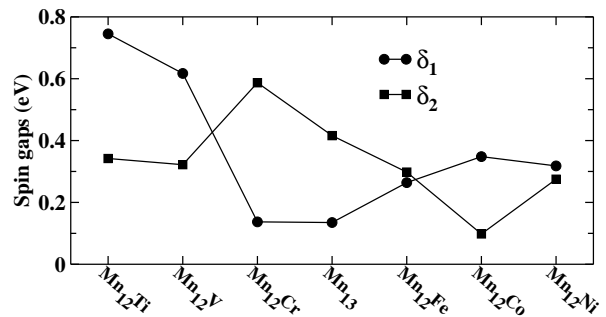}}
\caption{Plot of spin gaps [as defined in Eq. 2] of the MESs of Mn$_{12}$X clusters. }
\label{spingap2}
\end{center}
\end{figure}

To understand the relative stability of the doped Mn$_{12}$X clusters, we have computed the binding energies of MnX dimers as given in Table II and also plotted the spin gaps [as defined in Eq.\ref{spingap}] for the MESs of the Mn$_{12}$X clusters in Fig.\ref{spingap2}. Our calculated binding energies for dimers follow the same trend as we observed 
in case of clusters. It is seen from Table II that MnTi dimer has the maximum binding energy of 3.14 eV. Again the binding energy
 of MnV dimer ($\sim$ 2.72 eV) is also quite high. Therefore, the large gains in binding energies of Mn$_{12}$Ti and Mn$_{12}$V clusters come from the large binding energies of MnTi and MnV dimers respectively. On the other hand, the MnFe, MnCo and MnNi dimers also have large binding energies compared to that of the Mn$_2$ dimer. Consequently,
 the Mn$_{12}$Fe, Mn$_{12}$Co and Mn$_{12}$Ni clusters also have larger binding energies compared to the pure Mn$_{13}$ cluster.
\begin{table}[h]

\begin{center}
\caption{\label{table2}Binding energies and bond lengths of MnX dimers in the present calculation.
 For comparison, we have also listed the corresponding values as reported earlier (Ref. 26) for all the dimers.}
\vskip 0.2cm
\begin{tabular}{ccccccc}
\hline
\hline
Dimers    & & \multicolumn{2}{c} {Binding energy (eV)}& & \multicolumn{2}{c}{Bond length (\AA)} \\
\cline{3-4}\cline{6-7}
         & &       Present     &     Ref. 26                 & & Present       &   Ref. 26             \\
\hline
\\
MnTi     & &     3.14        &     2.76           & &  1.80         &  1.76  \\
MnV      & &     2.72        &     2.88           &  & 1.65         &  1.69  \\
MnCr     & &     1.17        &     0.93           & &  2.34         &  2.46  \\
Mn$_2$   & &     1.01        &     1.15           & &  2.58         &  2.62   \\
MnFe     & &     1.76        &     1.57           & &  2.39         &  2.42  \\
MnCo     & &     2.17        &     1.91           & &  2.06         &  2.09 \\
MnNi     & &     2.67        &     2.41           & &  2.06         &  2.09  \\
\hline
\end{tabular}
\end{center}
\end{table}

 For comparison, we have also listed the previously calculated values of binding energies and bond lengths of the dimers as reported in Ref.26. Our calculated values follow the similar trend as obtained from the reported values for both bondlenghts and binding energies of the dimers. However, we note that the reported binding energies in Ref. 26 show the least value for MnCr dimer and followed by Mn$_2$ dimer, while our calculated binding energies for dimers show the least value for Mn$_2$ dimer, followed by MnCr dimer. In order to verify our calculation for MnCr dimer, we have also performed the calculation for Cr$_2$ dimer.
 Our calculated binding energy and bond length for Cr$_2$ dimer are 2.08 eV and 1.59 {\AA} respectively. These values are consistent with the experimental values for Cr$_2$ dimer.\cite{cr2} Considering a mixed dimer with Mn and Cr, then the
 corresponding values for the MnCr dimer should be in between the corresponding values of the two pure dimers - Mn$_2$ and Cr$_2$. Our calculated values of binding energy and bond length for MnCr dimer follow this trend. 
Both the spin gaps ($\delta_1$ and $\delta_2$) as plotted in Fig.\ref{spingap2}, are also positive for all the clusters studied here indicating the formation of stable clusters. We find that the MESs of Mn$_{12}$Ti and Mn$_{12}$V clusters have large positive value of $\delta$'s, which in turn, again indicate the high stability of the Mn$_{12}$Ti and Mn$_{12}$V clusters compared to the others.

\begin{figure*}[]
\begin{center}
\rotatebox{0}{\includegraphics[height=4.0cm,keepaspectratio]{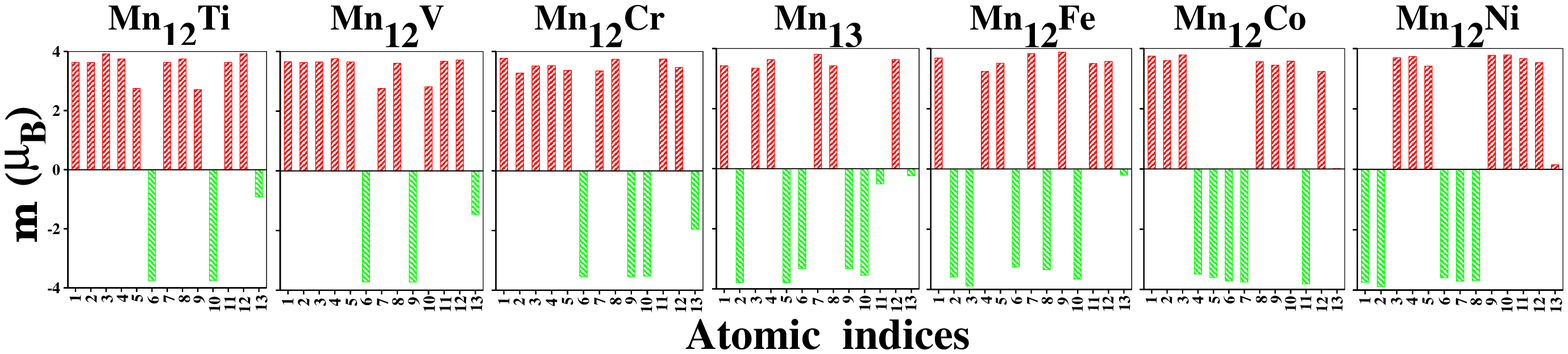}}
\caption{(Color online) Individual atomic magnetic moments ($m$) of the MESs of Mn$_{12}$Ti, Mn$_{12}$V, Mn$_{12}$Cr, Mn$_{13}$, Mn$_{12}$Fe, Mn$_{12}$Co and Mn$_{12}$Ni clusters. The atomic indices 1-12 correspond to the surface atoms (same for each Mn$_{12}$X cluster), while index `13' corresponds to the central atom, as shown in Fig.2. Lengths of the bars correspond to the magnitude of atomic moments. Red colored (dark) bars in the positive direction correspond to +ve magnetic moments and green colored (gray) bars in the negative direction for -ve magnetic moments. The central site is the most coordinated site and therefore has very small magnetic moment for all the clusters. }
\label{spin_barplot}
\end{center}
\end{figure*}

To understand the magnetic behavior of the MESs of the Mn$_{12}$X clusters, we have calculated the atom-centered magnetic moments within a specified sphere around each atom of the MESs of doped Mn$_{12}$X clusters and pure Mn$_{13}$ cluster. Fig. \ref{spin_barplot} shows the pictorial representation of our analysis. Focusing first on the individual moments, we find that the moments of all the atoms for each cluster remain close to the atomic moment ($\sim$ 3-4 $\mu_B$) with the exception of the highly coordinated central atom (atomic index 13) which has a small magnitude for all the Mn$_{12}$X clusters. It is therefore the alignment of individual moments of the surface atoms which plays the major role in determining the total magnetic moment of the MESs of the Mn$_{12}$X clusters. In case of doping with atom of elements left of Mn in the periodic table (i.e for the Mn$_{12}$Ti, Mn$_{12}$V and Mn$_{12}$Cr clusters), it is seen that the surface Mn-atoms  are mostly ferromagnetically aligned. For example, only the moments of two surface atoms are antiferromagnetically aligned with the moments of other surface atoms in case of MESs for Mn$_{12}$Ti and Mn$_{12}$V clusters. Similarly, moments of three surface Mn atoms are antiferromagnetically aligned with the moments of other surface atoms in case of the MES for Mn$_{12}$Cr cluster. On the other hand, the surface Mn atoms are mostly antiferromagnetically coupled among themselves in case of  substitution with atom of the elements right of Mn in the periodic table (i.e for Mn$_{12}$Fe, Mn$_{12}$Co and Mn$_{12}$Ni clusters) and the net magnetic moment for each of these clusters is small, close to the total magnetic moment of two ferromagnetically coupled Mn atoms which is 10 $\mu_B$.\cite{ref4}

\begin{figure}[]
\begin{center}
\vskip 1.7cm
\rotatebox{0}{\includegraphics[height=4.9cm,keepaspectratio]{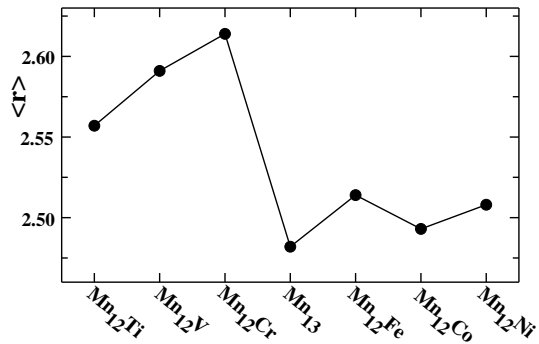}}
\caption{Plot of centre to surface average bondlengths ($\langle$r$\rangle$) of the MESs of Mn$_{12}$X clusters. }
\label{bondlength}
\end{center}
\end{figure}

 We further find that the magnetic moment of the central atom, though it is small compared to that of the surface atoms for each doped cluster, varies  from one doped cluster to another. It shows a systematic increase as one moves from Mn$_{12}$Ti$\rightarrow$Mn$_{12}$V$\rightarrow$Mn$_{12}$Cr and attains tiny values for Mn$_{13}$, Mn$_{12}$Fe, Mn$_{12}$Co and Mn$_{12}$Ni. The moment at the central site is decided by two factors - (i) gradual filling of electronic states in moving from left to right of the periodic table and (ii) the hybridization with neighboring atoms. The gradual filling of electronic states should increase the moment in moving from Ti to V to Cr attaining a maximum value for the half-filled case of Mn, followed by a gradual decrease in moving from Fe to Co to Ni. While the above mentioned trend is observed for the central atom of Mn$_{12}$Ti, Mn$_{12}$V and Mn$_{12}$Cr clusters, the moments for the central atom are all tiny for Mn$_{13}$, Mn$_{12}$Fe, Mn$_{12}$Co and Mn$_{12}$Ni clusters, in contrast to the electron filling trend which would demand the moment to be the maximum for Mn$_{13}$. This is, however, explained by the second effect, namely the hybridization effect. As the bond length decreases, generally the magnetic moment also decreases through the increase in effective hybridization. In order to examine the effect of hybridization, we have, therefore, calculated the centre to surface average distance for the MESs of all the seven Mn$_{12}$X clusters with X = Ti, V, Cr, Mn, Fe, Co, Ni. Fig. \ref{bondlength} shows the plot of our calculation. We find the average bond lengths to grow in moving from Mn$_{12}$Ti to Mn$_{12}$V to Mn$_{12}$Cr and then dropping suddenly to a smaller value for Mn$_{13}$ which stays as smaller value for Mn$_{12}$Fe, Mn$_{12}$Co and Mn$_{12}$Ni clusters also. The hybridization effect therefore takes over the filling effect in producing small magnetic moment at the central sites for Mn$_{13}$ as well as Mn$_{12}$Fe, Mn$_{12}$Co and Mn$_{12}$Ni. Therefore, we find that the central atom behaves as magnetic atom in case of Mn$_{12}$Ti, Mn$_{12}$V and Mn$_{12}$Cr clusters, while it behaves as essentially nonmagnetic atom in case of Mn$_{13}$, Mn$_{12}$Fe, Mn$_{12}$Co and Mn$_{12}$Ni clusters. Considering the alignment of moments for both the surface atoms as well as central atom, we can then conclude that when the central atom is magnetic, the surface Mn-atoms are mostly antiferromagnetically coupled with the central atom, causing the effective ferromagnetic coupling among the surface atoms which gives rise to large total magnetic moment of the clusters. On the other hand, when the central atom is nonmagnetic, the surface magnetic moments interact with each other through super-exchange path involving the central nonmagnetic atom, resulting into antiferromagnetic coupling within each other. This gives rise to a small net magnetic moment of the whole clusters.

\subsection{N-capped Mn$_{13}$N cluster}

\begin{figure}[h]
\begin{center}
\vskip 0.3 cm
\rotatebox{0}{\includegraphics[height=5.9cm,keepaspectratio]{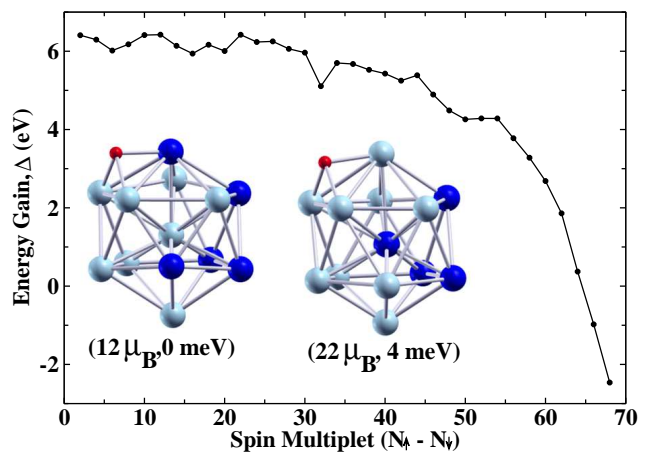}}
\caption{(Color online) Variation of energy gain $\Delta$ as defined in the text as a function of total magnetic moments of the Mn$_{13}$N cluster. The insets show the atomic spin orientation of the minimum energy icosahedral structure and the first isomer of Mn$_{13}$N cluster. The cyan (light gray) colored balls correspond to atoms with positive magnetic moment and blue (dark gray) colored balls correspond to atoms with negative magnetic moment The N-atom [represented by the smaller sized red (dark gray) colored ball] in case of both the structures of inset, have negative magnetic moment. The numbers in the parentheses for the insets correspond to total magnetic moment and the relative energy to the MES of Mn$_{13}$N cluster respectively.}
\label{figMn13N}
\end{center}
\end{figure}

Finally, in order to see the effect of N-capping in Mn$_{13}$ cluster along the same direction of Ref.12, we have also studied the structure, stability and magnetic properties 
of an icosahedral Mn$_{13}$N cluster with N atom at a triangular face cap as shown in the inset of Fig.\ref{figMn13N}. To determine the magnetic moment of the MES and to identify the possible
 isomers close to MES, we have again optimized the structure for {\it all possible} collinear spin configurations of atoms. We find an icosahedral structure of total magnetic moment of 12 $\mu_B$ is the MES for Mn$_{13}$N cluster.  We also find that the Mn-N-Mn bond angles of N atom with the three nearest Mn-atoms in case of the MES of Mn$_{13}$N cluster, are 90.08$\degree$, 93.95$\degree$ and 90.97$\degree$. This nearly 90$\degree$ bond angle between Mn-N-Mn has been predicted as guiding rule for the ground state structure of N-caped Mn-clusters in the earlier work.\cite{ref5} It results from the interaction of the $p$-orbitals of N-atom with the $s$-$d$ hybridized orbitals of Mn atoms. As the N atom is added, the binding energy of  Mn$_{13}$N cluster improves significantly compared to that of the MES for pure Mn$_{13}$ cluster. The energy gained in adding a N atom to Mn$_{13}$ cluster, is defined as $\Delta = -[E(Mn_{13}N)-E(Mn_{13})-E(N)]$. 

Fig.\ref{figMn13N} shows the variation of $\Delta$ as a function of total magnetic moment of the Mn$_{13}$N cluster. The MES of Mn$_{13}$N cluster has binding energy gain $\Delta$ = 6.42 eV. There are also several closely lying isomers with total magnetic moments of 22 $\mu_B$, 10 $\mu_B$ and 2 $\mu_B$ which are 4 meV, 10 meV and 14 meV away from that of the MES respectively. So $\Delta$'s for them are also very close to that of the MES. These values of $\Delta$'s are consistent with the energy-gain predicted in Ref.12 for Mn$_5$ cluster, which is 5.61 eV. In the earlier work, \cite{ref5} it has been shown that addition of one N-atom to at least five Mn atoms is energetically favorable. Here we see that addition of single N atom is also energetically favorable even for the Mn$_{13}$ cluster. This enhancement in bonding due to N-addition has been demonstrated earlier\cite{ref5}
 by examining the bonding in case of dimer - Mn$_2$ versus Mn$_2$N. Lack of hybridization between the $s$ and $d$ electrons due to filled 4$s$ shell of Mn prevents Mn$_2$ from forming 
a strong bond. As N is attached, the 4$s^2$ electrons of Mn interact with the 2p$^3$ electrons of N resulting in strong coupling. The inset of Fig.\ref{figMn13N} shows the orientations of atomic spins of the minimum energy structure of magnetic moment 12 $\mu_B$ and the first isomer of magnetic moment 22 $\mu_B$ of Mn$_{13}$N cluster. The local effect of N-capping in now very obvious. Effectively, it couples the atomic spins ferromagnetically in the triangular face where it sits as well as in the pentagonal ring nearest to it (upper pentagonal rings of the two structures in the insets of Fig.\ref{figMn13N}). Therefore, two pentagons for each of the two structures (MES and the closely lying first isomer) in the inset of Fig.\ref{figMn13N}, are now effectively ferromagnetically coupled. So in addition of enhancing binding, N-capping also increases the magnetic moment compared to that of the pure Mn$_{13}$ cluster. However, it is significantly lower than that of the MESs for Mn$_{12}$Ti and Mn$_{12}$V clusters, presumably due to the fact that N-capping produces a local effect while the Ti or V substitutions at the central site effects all the other 12 surface atoms thereby producing a global effect.

\section{Summery and Conclusions}

Total magnetic moment of a Mn$_{13}$ cluster is small due to anti-ferromagnetic alignment of individual atomic moments. Our study shows that it is possible to obtain stable ferromagnetic alignment of atomic spins by substitutional Ti-doping and V-doping in Mn$_{13}$ cluster resulting into giant total magnetic moment. We have compared our findings with the previously proposed route of $N$-capping to Mn-clusters. The gain in magnetic moment compared to pure Mn$_{13}$ cluster is found to be more in case of Ti and V substitutions.
 Also substitutional doping keeps the volume of the parent cluster intact, while the N-capping increases the effective volume of the cluster. Therefore, we propose Ti-substitution and V-substitution as promising alternative to engineer the magnetic structure for Mn$_{13}$ cluster.

\acknowledgments
 T.S.D., A.M. and S. D. thank Department of Science and Technology, India for the support through
Advanced Materials Research Unit.

\end{document}